\documentclass[twocolumn,aps,showpacs,prl]{revtex4-1}

\usepackage{amsmath,amssymb,graphics,epsfig,epstopdf,color,multirow,array,verbatim,ulem,braket,tabularx}
\usepackage[colorlinks,linkcolor=blue,citecolor=blue,urlcolor=blue]{hyperref}

\begin{document}

\title{LnPd$_{2}$Sn (Ln=Sc, Y, Lu) class of Heusler alloys for topological superconductivity }

\author{Peng-Jie Guo$^{1}$}
\author{Jian-Feng Zhang$^{1}$}
\author{Huan-Cheng Yang$^{2,1}$}
\author{Zheng-Xin Liu$^{1}$}
\author{Kai Liu$^{1}$}\email{kliu@ruc.edu.cn}
\author{Zhong-Yi Lu$^{1}$}\email{zlu@ruc.edu.cn}

\affiliation{$^{1}$Department of Physics and Beijing Key Laboratory of Opto-electronic Functional Materials $\&$ Micro-nano Devices, Renmin University of China, Beijing 100872, China}
\affiliation{$^{2}$Beijing Computational Science Research Center, Beijing 100193, China}

\date{\today}

\begin{abstract}
 Based on the first-principles electronic structure calculations and the symmetry analysis, we predict that the topological superconductivity may occur on the surface of the LnPd$_{2}$Sn (Ln=Sc, Y, Lu) class of Heusler alloys. The calculated electronic band structure and topological invariant demonstrate that the LnPd$_{2}$Sn family is topologically nontrivial. The further slab calculations show that the nontrivial topological surface states of LnPd$_{2}$Sn exist within the bulk band gap and meanwhile they cross the Fermi level. Considering that the LnPd$_{2}$Sn class of compounds were all found experimentally to be superconducting at low temperature, the surface topological superconductivity is likely to be generated via the proximity effect. Thus the LnPd$_{2}$Sn class of compounds shall be a promising platform for exploring novel topological superconductivity and handling Majorana zero modes.
\end{abstract}

\date{\today} \maketitle



Topological superconductors that hold Majorana zero modes (MZM) at open boundaries, vortex cores or other topological defects have attracted lots of interest in condensed matter physics in recent years. In contrast to the Majorana fermions which are propagating particles in high energy physics, MZM are bound states which give rise to degeneracy of the ground states and obey non-Abelian braiding statistics if they are adiabatically exchanged in space. Among all the possible mechanisms that host non-Abelian anyon excitations, topological superconductors are most promising to be realized experimentally and have great potential applications in fault-tolerent topological quantum computations~\cite{rmp2008,rmp2011,rpp2017}. To well study the novel properties of MZM, it is necessary to explore a variety of topological superconductors as many as possible, especially those with high-quality samples. Unfortunately, to date, intrinsic spinless $p$+i$p$ type topological superconductors are very rare \cite{sro2003rmp}. Accordingly, an approach to realize the equivalent $p$+i$p$ type topological superconductivity has been proposed, namely  to construct a heterostructure between a topological insulator and an $s$-wave superconductor (TI/SC)~\cite{fu2008prl}, at whose interface the topological superconductivity is induced via the proximity effect and the MZM are reported to be observed at the vortices by scanning tunneling microscope (STM)~\cite{science2012, science2014, nature2016, np2014, sunprl}. Such a heterostructure, however, is confronted with the complex interface effect. An alternative approach is to induce superconductivity in a doped topological insulator, which is confronted with the disorder effect due to doping~\cite{hnp}. To avoid these challenges, a natural idea is to search for the equivalent $p$+i$p$ type topological superconductivity in a single compound without chemical doping or causing interfaces.

Compared with the heterostructure of TI/SC or the doped topological insulators, such a single-compound has to meet the following four strict requirements simultaneously: (i) The compound is a topological insulator defined on a curved Fermi surface; (ii) The compound is an $s$-wave superconductor; (iii) The nontrivial topological surface states of the compound are not overlapped with the bulk states; (iv) These nontrivial topological surface states cross the Fermi level. It turns out that such compounds are also very rare to date~\cite{M-nc, a15arxiv, zhangarxiv}.

The Heusler alloys are very abundant in nature and own a variety of novel physical properties~\cite{graf2011}, covering magnetism~\cite{prl1983,prb1983,prl1997}, superconductivity~\cite{ssc1984, ml1983, prb1985}, as well as nontrivial topology~\cite{snm2010, hnm2010, prb2017, prl2017}. Thus, there is a strong expectation and demand that we may realize topological superconductivity in the family of Heusler alloys. In this Letter, based on the first-principles electronic structure calculations and the symmetry analysis, we present the nontrivial topological properties of LuPd$_{2}$Sn as a representative of the LnPd$_{2}$Sn class of Heusler alloys, whose nontrivial surface states are not overlapped with the bulk states and meanwhile they cross the Fermi level. Since these compounds were found experimentally to be superconducting at low temperature, they are a promising class of single-compound materials for implementing surface topological superconductivity.


The electronic structures of LnPd$_{2}$Sn (Ln=Sc, Y, Lu) were studied based on the first-principles electronic structure calculations. The projector augmented wave (PAW) method \cite{pg-prb} as implemented in the VASP package \cite{gg-3} was used to describe the core electrons. The generalized gradient approximation (GGA) of Perdew-Burke-Ernzernof (PBE) type was adopted for the exchange-correlation functional. The kinetic energy cutoff of the plane wave basis was set to 400 eV. A 10$\times$10$\times$10 $k$-point mesh for the Brillouin zone (BZ) sampling and the Gaussian smearing method with a width of 0.05 eV around the Fermi surface were utilized. Both the cell parameters and the internal atomic positions were fully relaxed until the forces on all atoms were smaller than 0.01 eV/{\AA}. Once the equilibrium structures were obtained, the electronic structures were further calculated by including the spin-orbit coupling (SOC). In order to study the topological surface states of LnPd$_{2}$Sn (Ln=Sc, Y, Lu), a two-dimensional supercell with a 58-atom slab and a 20 {\AA} vacuum were employed.


The full Heusler alloys LnPd$_{2}$Sn (Ln=Sc, Y, Lu) own a symmorphic space group symmetry $Fm\bar3m$. From the crystal structure shown in Fig. 1(a), we can see that the Ln (Ln=Sc, Y, Lu) and Sn atoms form a rocksalt structure, while the Pd atoms locate at the equivalent (1/4, 1/4, 1/4) fractional coordinates. The corresponding bulk BZ along with the the high-symmetry $k$ points is demonstrated in the bottom part of Fig. 1(b). The projected two-dimensional (2D) BZ of the (001) surface and the corresponding high-symmetry $k$ points in 2D BZ are shown in the upper part of Fig. 1(b).

\begin{figure}[!t]
\includegraphics[width=1\columnwidth]{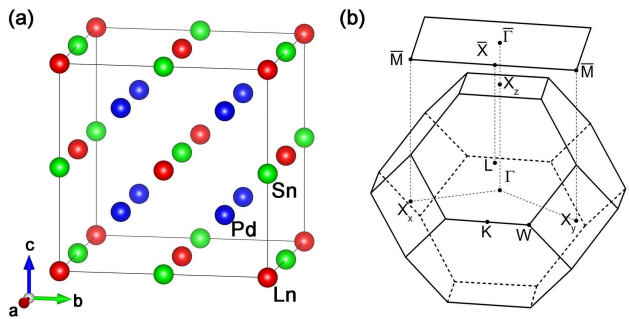}
\caption{(Color online) (a) The crystal structure of full Heusler alloys LnPd$_{2}$Sn (Ln=Sc, Y, Lu). (b) The bulk Brillouin zone (BZ) and the projected two-dimensional (2D) BZ of the (001) surface. Here the black dots represent the high-symmetry $k$ points in BZ.}
\label{Fig1}
\end{figure}

As the LnPd$_{2}$Sn (Ln=Sc, Y, Lu) class of compounds have similar electronic band structures, here we choose LuPd$_{2}$Sn as a representative one. Figures 2(a) and 2(b) show the electronic band structures of LuPd$_{2}$Sn calculated without and with the SOC, respectively. There are three bands crossing the Fermi level, resulting in three hole-type Fermi surfaces. For the results without the SOC, the valence band (in red color) and the conduction band (in blue color) have a crossing along the X-$\Gamma$ path in the BZ. Based on the symmetry analysis, these two crossing bands belong to the $\triangle_{2}$ and $\triangle_{4}$ irreducible representations of the C$_{4v}$ group [Fig. 2(a)], respectively. Thus, the crossing point along the X-$\Gamma$ path located about 0.2 eV above the Fermi level ($E_F$) is a Dirac point, which is protected by the C$_{4v}$ group symmetry and the SU(2) spin rotation symmetry.

\begin{figure}[!t]
\includegraphics[width=0.92\columnwidth]{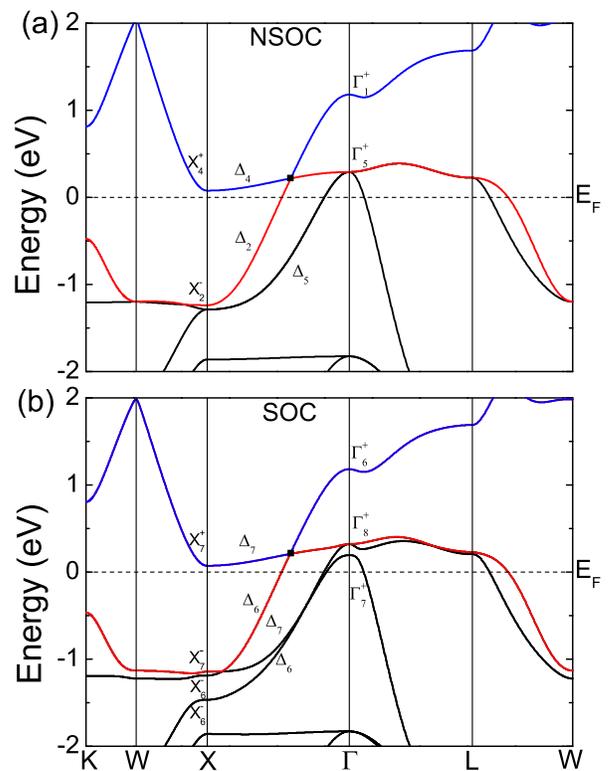}
\caption{(Color online) Electronic band structures of LuPd$_{2}$Sn calculated (a) without and (b) with the spin-orbit coupling (SOC) along the high-symmetry paths of the bulk BZ. The symmetries of these bands close to the Fermi level are also labeled. Here the black squares denote the Dirac points.}
\label{Fig2}
\end{figure}

With inclusion of the SOC, the symmetry of LuPd$_{2}$Sn can be described by the corresponding double group. According to the group theory calculations, the $\triangle_{2}$ and $\triangle_{4}$ irreducible representations of the C$_{4v}$ group change to the $\triangle_{6}$ and $\triangle_{7}$ irreducible representations of the C$_{4v}$ double group [Fig. 2(b)], respectively. As a result, the Dirac point along the X-$\Gamma$ path located 0.2 eV above $E_F$ is still stable when including the SOC. On the other hand, there is another band-crossing along the X-$\Gamma$ path located about 1.1 eV below $E_F$ with the inclusion of the SOC, for which the two crossing bands belong respectively to the $\triangle_{6}$ and $\triangle_{7}$ irreducible representations of the C$_{4v}$ double group [Fig. 2(b)]. In other words, this crossing point is also a Dirac point protected by the C$_{4v}$ double group symmetry. Moreover, the group theory calculations demonstrate that the $\triangle_{5}$ irreducible representation of the C$_{4v}$ group splits into a direct sum of the $\triangle_{6}$ and $\triangle_{7}$ irreducible representations of the C$_{4v}$ double group when including the SOC. Thus, if two doubly degenerated bands belonging to the $\triangle_{5}$ irreducible representation of the C$_{4v}$ group in the absence of the SOC generate a crossing point when turning on the SOC, the crossing point is a Dirac point protected by the corresponding C$_{4v}$ double group symmetry. Interestingly, the two crossing bands along the X-$\Gamma$ path in black color [Fig. 2(b)] completely satisfy the above condition. Besides the Dirac points along the X-$\Gamma$ path, the red band and the black band derived from a doubly degenerated band without the SOC [Fig. 2(a)] have also a crossing point along the $\Gamma$-L path of the BZ [Fig. 2(b)]. According to the group theory calculations, the two-dimensional irreducible representation of the C$_{3v}$ group along the $\Gamma$-L path splits into one two-dimensional and two one-dimensional irreducible complex representations of the corresponding C$_{3v}$ double group. As there are both time-reversal and space-inversion symmetries, the two one-dimensional irreducible representations must be degenerate to form a Kramer's doublet. Thus, this crossing point along the $\Gamma$-L path is also a Dirac point protected by the C$_{3v}$ double group symmetry. Since there are three equivalent X points and four equivalent L points in the BZ [Fig. 1(b)] and the corresponding Dirac cones are all strongly tilted [Fig. 2(b)], in total LuPd$_{2}$Sn has 13 pairs of type-II Dirac points in the energy ranging from -1.5 to 1.0 eV around the Fermi level. Thus LuPd$_2$Sn is a type-II Dirac semimetal.

\begin{table}[!b]
\caption{\label{tab:I} The calculated parity products of all occupied bands at eight time-reversal invariant momentum (TRIM) points in the BZs of ScPd$_{2}$Sn, YPd$_{2}$Sn, and LuPd$_{2}$Sn. The corresponding superconducting transition temperatures $T_c$ reported in literatures are also listed.}
\begin{center}
\begin{tabular*}{0.95\columnwidth}{@{\extracolsep{\fill}}ccccc}
\hline\hline
 &   $\Gamma$  &  3X  &  4L  &  $T_{c}$ \\
\hline
 ScPd$_{2}$Sn & + & - & + &  2.25$^a$ \\
 YPd$_{2}$Sn  & + & - & + &  4.9$^b$ \\
 LuPd$_{2}$Sn & - & + & + &  3.05$^c$ \\
\hline\hline
$^{a}$Ref.~\onlinecite{ssc1984}.\\
$^{b}$Ref.~\onlinecite{ml1983}.\\
$^{c}$Ref.~\onlinecite{prb1985}.
\end{tabular*}
\end{center}
\end{table}

\begin{figure}[!t]
\includegraphics[width=0.92\columnwidth]{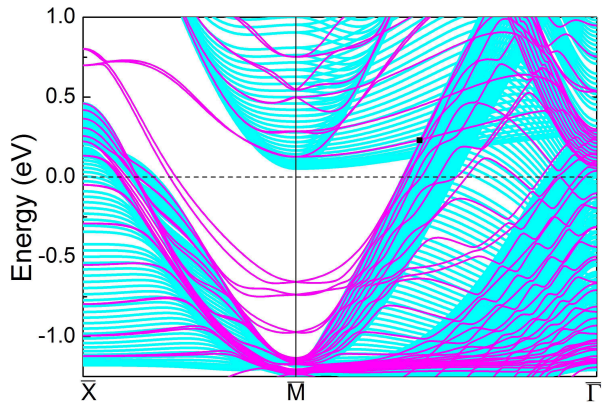}
\caption{(Color online) Surface band structure of the Pd-terminated LuPd$_{2}$Sn(001) surface along the high-symmetry paths of the surface BZ [Fig. 1(b)]. The cyan bands are the surface-projected bulk states. Here the Fermi level sets to zero and the black square denotes the projected bulk Dirac point.}
\label{Fig3}
\end{figure}

Nevertheless, here more importantly, the parities of the two bands in red and blue colors around the $E_F$ at the high-symmetry X point are inversed for LuPd$_{2}$Sn (Fig. 2). As there are odd numbers of X points in the BZ, these band inversions may lead to nontrivial topological properties. When a crystal owns inversion symmetry, its Z$_{2}$ topological invariant can be calculated by the product of the parities of all occupied states at all the time-reversal invariant momentum (TRIM) points \cite{fu-prl}. Although the red-color band and the blue-color band of LuPd$_{2}$Sn have contact points along the X-$\Gamma$ paths, we can still calculate the Z$_{2}$ topological invariant by this method, for which the highest occupied band is the one in red color. The calculations show that the Z$_{2}$ topological invariant of LuPd$_{2}$Sn equals to 1 (Table I), indicating that LuPd$_{2}$Sn can be considered as a strong topological insulator defined on a curved Fermi surface between the blue-color band and the red-color band [Fig. 2(b)]. Likewise, ScPd$_{2}$Sn and YPd$_{2}$Sn are type-II Dirac semimetals and can be also considered as strong topological insulators defined on a similar curved Fermi surface (Table I).

\begin{figure}[!t]
\includegraphics[width=0.95\columnwidth]{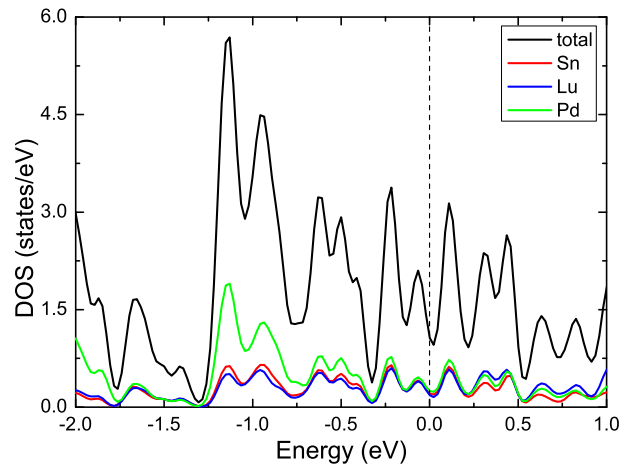}
\caption{(Color online) Total and local density of states for LuPd$_{2}$Sn calculated with the SOC. Here the Fermi level sets to zero.}
\label{Fig4}
\end{figure}

The nontrivial topology of the band structure is always accompanied by the nontrivial topological surface states, which is the most important phenomenon for the topological materials. We have thus studied the surface states of LuPd$_{2}$Sn(001) surface. For this surface, the X$_{z}$ point in the bulk BZ is projected to the $\bar{\Gamma}$ point in the 2D BZ, while both the X$_{x}$ and X$_{y}$ points are projected to the $\bar{M}$ points [Fig. 1(b)]. The band structure of Pd-terminated LuPd$_{2}$Sn(001) surface is demonstrated in Fig. 3. Compared with the bulk states (in cyan color), there are the surface states (in magenta color) crossing the Fermi level and merging into the bulk states at the projected bulk Dirac point along the $\bar{M}$-$\bar{\Gamma}$ path denoted by the black square in Fig. 3, corresponding to the bulk Dirac point along the X-$\Gamma$ path of the bulk BZ denoted by the black square in Fig. 2(b). These surface states are thus nontrivial topological surface states and are protected by the time-reversal symmetry. Notably, these surface states are not overlapped with the bulk states and can be readily observed by angle-resolved photoemission spectroscopy (ARPES) experiment. We have also calculated the surface states of ScPd$_{2}$Sn and YPd$_{2}$Sn, and found that they show the similar characteristics.

We have further studied the electronic states of LuPd$_{2}$Sn. As shown in Fig. 2, there are two hole-type bands around $E_F$ that are relatively flat along the X-$\Gamma$ and $\Gamma$-L paths. This may induce a large density of states (DOS) around the Fermi level, which is verified by the DOS calculations in Fig. 4. As further indicated by the local density of states, the three atomic species (Lu, Pd, Sn) have almost the same contributions around the $E_F$. It is well known that the large DOS close to the Fermi level is beneficial for superconductivity. Actually, previous transport experiments demonstrated that ScPd$_{2}$Sn, YPd$_{2}$Sn, and LuPd$_{2}$Sn are all superconducting with respective transition temperatures $T_c$ of 2.25 K \cite{ssc1984}, 4.9 K \cite{ml1983}, and 3.05 K \cite{prb1985} (Table I).


So far, we have demonstrated that LnPd$_{2}$Sn (Ln=Sc, Y, Lu) are type-II Dirac semimetals and meanwhile can be also considered as topological insulators defined on a curved Fermi surface. Here we would like to emphasize that the nontrivial topological surface states of LnPd$_{2}$Sn at the (001) surface cross the Fermi level and isolate from the bulk states. Moreover, LnPd$_{2}$Sn are all superconducting at low temperatures \cite{ssc1984, ml1983, prb1985}. Thus, via the proximity effect, the topological superconductivity in the nontrivial surface states of LnPd$_{2}$Sn may be induced by the bulk superconductivity~\cite{fu2008prl,M-nc}.

As a new class of materials for topological superconductivity, LnPd$_{2}$Sn (Ln=Sc, Y, Lu) have many advantages. Firstly, due to the cubic symmetry, the (100) and (010) surfaces of LnPd$_{2}$Sn possess the same nontrivial topological surface states and possible topological superconductivity as the (001) surface, which can be observed by the ARPES experiment. Secondly, compared with the doped topological insulators or the TI/SC heterostructures, the LnPd$_{2}$Sn class of single-compound materials avoid the disorder effect as well as the complex interface effect, but meanwhile very likely maintain the robustness of Majorana zero modes against the other bound states in vortex cores, namely thermal effects, by the enlarged minigaps due to the proximity-effect \cite{sau}. This will be helpful to study the exotic properties of topological superconductivity and Majorana zero modes in experiment. Last but not least, so far the single-compound topological superconductors without doping are very few, only three candidates, namely $\beta$-PdBi$_{2}$ \cite{M-nc}, Ta$_{3}$Sb \cite{a15arxiv} and $\beta$-RhPb$_{2}$ \cite{zhangarxiv}. Here in the LnPd$_{2}$Sn class, there are also three candidates, especially, the superconducting transition temperature $T_c$ of YPd$_{2}$Sn is 4.9 K \cite{ml1983}, right above the liquid-helium temperature.


In summary, by using the symmetry analysis and the first-principles electronic structure calculations, we have systematically studied the topological properties of the LnPd$_{2}$Sn (Ln=Sc, Y, Lu) class of Heusler alloys. Notably, the nontrivial topological surface states of LnPd$_{2}$Sn are not overlapped with the bulk states and meanwhile they cross the Fermi level. In combination with the fact that the LnPd$_{2}$Sn (Ln=Sc, Y, Lu) compounds are all superconducting at low temperatures, their bulk superconductivity may induce the superconductivity in the nontrivial topological surface states via the proximity effect. Furthermore, due to the cubic symmetry of these compounds, the equivalent (100), (010), and (001) surfaces may all hold topological superconductivity. Thus, the LnPd$_{2}$Sn (Ln=Sc, Y, Lu) class of compounds may be a promising platform for studying the novel properties of the topological superconductor and handling Majorana zero modes in future experiments.

\begin{acknowledgments}

We wish to thank Xin Liu for helpful conversations. This work was supported by the National Key R\&D Program of China (Grants No. 2017YFA0302903 and No. 2016YFA0300504), the National Natural Science Foundation of China (Grants No. 11774422, No. 11774424, and No. 11574392). Computational resources were provided by the Physical Laboratory of High Performance Computing at Renmin University of China.

\end{acknowledgments}

\end{document}